\documentclass[12pt]{article}

\usepackage{graphicx}
\usepackage{amssymb}
\usepackage{amsmath}
\usepackage{color}

\def\bea{\begin{eqnarray}}
\def\eea{\end{eqnarray}}
\def\beq{\begin{eqnarray}}
\def\eeq{\end{eqnarray}}

\def\bm{\begin{math}}
\def\me{\end{math}}

\begin{document}

\begin{center}
{\Large{\bf Surface-Directed Spinodal Decomposition: A Molecular Dynamics Study}} \\
\ \\
\ \\
by \\
Prabhat K. Jaiswal$^1$, Sanjay Puri$^1$, and Subir K. Das$^{2}$ \\
$^1$School of Physical Sciences, Jawaharlal Nehru University, New Delhi -- 110067, India. \\
$^2$Theoretical Sciences Unit, Jawaharlal Nehru Centre for Advanced Scientific Research, 
Jakkur, Bangalore -- 560064, India.
\end{center}

\begin{abstract}
We use molecular dynamics (MD) simulations to study
surface-directed spinodal decomposition (SDSD) in unstable
binary ($AB$) fluid mixtures at wetting surfaces.
The thickness of the wetting layer $R_1$ grows with time $t$ as
a power-law ($R_1 \sim t^\theta$). We find that hydrodynamic
effects result in a crossover of the growth exponent from
$\theta\simeq 1/3$ to $\theta\simeq1$. We also present results for the
layer-wise correlation functions and domain length scales.
\end{abstract}

\newpage                             
\section{Introduction}
\label{sec:intro}

There has been great interest in problems of {\it phase ordering dynamics} in 
recent years. A prototypical problem in this area is the phase-separation kinetics 
of a homogeneous binary ($AB$) mixture which has been rendered thermodynamically unstable by a rapid quench below the miscibility curve. If the quenched mixture is spontaneously unstable, the evolution 
kinetics is usually referred to as {\it spinodal decomposition} (SD). 
During SD, there is emergence and growth of $A$-rich and  $B$-rich domains, characterized by a single 
time-dependent length scale $L(t)$. This has important consequences, e.g., the correlation function of the order parameter field exhibits the scaling form $C(r,t)=f[r/L(t)]$, where $f(x)$ is a scaling function. We now have a good 
understanding of the kinetics of phase separation in the bulk, and there are several good reviews of these problems 
\cite{pw09,bray94,bf01,onuki02}. 

Next, let us consider the equilibrium behavior of an immiscible $AB$ mixture in contact with a surface $S$.
Typically, the surface has a preferential attraction for one of the components of the 
mixture, say $A$. Let $\gamma_A$ and $\gamma_B$ be the surface tensions between the 
$A$-rich and $B$-rich phases and $S$, respectively, and let $\sigma$ be the surface 
tension between the $A$-rich and $B$-rich phases. We focus on a semi-infinite geometry 
for simplicity. Then the contact angle $\theta$ between the $AB$ interface 
and the surface can be obtained from Young's equation \cite{young}:
\begin{equation}
 \sigma\,\mathrm{cos}\theta = \gamma_{B} - \gamma_{A}.
 \label{eq:young}
\end{equation}
When $\gamma_{B}-\gamma_{A} > \sigma$, the $A$-rich phase covers the 
surface in a {\it completely wet} (CW) 
morphology. However, for $\gamma_{B}-\gamma_{A} < \sigma$, 
both phases are in contact with the 
surface resulting in a {\it partially wet} (PW) equilibrium morphology.

We have a long-standing interest in the kinetics of binary mixtures at surfaces.
Consider a homogeneous $AB$ mixture at high temperatures. This mixture is 
kept in contact with a surface which prefers $A$. The system is quenched deep below the 
miscibility curve at time $t=0$. Then, the system becomes unstable to phase separation and 
decomposes into $A$-rich and $B$-rich domains. The surface is simultaneously wetted 
by $A$. The interplay of these two dynamical processes, i.e., wetting and phase 
separation, is referred to as {\it surface-directed spinodal decomposition} (SDSD) or 
{\it surface-directed phase separation} 
\cite{jnkbw91,kdkb93,lwll09,wcl10,gk03,pb92,pb94,bc92,jm93,pbf97,pf97,pb01,puri05,dphb05,dphb06,yx08,ylx08,bpdh10}. 
These processes have important technological applications, including the fabrication of nanoscale patterns and multi-layered structures. 

With some exceptions \cite{bpl01,tanaka01}, most available studies of SDSD do not take into account hydrodynamic effects, i.e., the growth of bulk domains and the wetting layer is governed by diffusion. However, 
many important experiments in this area involve fluid or polymer mixtures, where fluid velocity fields play a substantial role in determining physical properties. Hydrodynamic effects alter the late-stage dynamics of phase separation in a drastic manner -- both without surfaces \cite{pw09,bray94,kcpdb01,wc01,adp10} and with surfaces 
\cite{bpl01,tanaka01}. In this paper, we have undertaken extensive molecular dynamics (MD) simulations to investigate the effects of hydrodynamics on the late-stage dynamics of SDSD. A preliminary account of our results was published as a recent letter \cite{jpd12}. We observe a clear crossover from a diffusive regime to a hydrodynamic regime in the growth law for the wetting layer. 
 
This paper is structured as follows: In Sec.~\ref{sec:md}, we describe the details of our MD simulations. Section~\ref{sec:theory} presents a brief review of bulk phase-separation kinetics and domain growth laws, and then discusses phase separation at surfaces. Detailed MD results are presented in Sec.~\ref{sec:results}. We end with a summary and discussion of our results in Sec.~\ref{sec:sum}.

\section{Details of Simulations}
\label{sec:md}

We employ standard MD techniques for our simulations \cite{at87,fs02}. The model is similar to that used in our earlier studies of mixtures at surfaces \cite{dphb06, jpd10}. We consider a binary fluid mixture $AB$
consisting of $N_A$ $A$-atoms and $N_B$ $B$-atoms (with $N_A=N_B$), confined in a box 
of volume $L_w \times L_w \times D$. While periodic boundary conditions are maintained 
in the $x$- and $y$-directions, walls or surfaces are introduced in the $z$-direction
at $z=0$ and $z=D$. The interaction between two atoms of species $i$ and 
$j$ separated by a distance $r$ is given by the Lennard-Jones (LJ) potential:
\begin{equation}
 u_{ij}(r)=4\epsilon_{ij}\left[\left(\frac{\sigma}
{r}\right)^{12}-\left(\frac{\sigma}{r}\right)^{6}\right]; \quad i, j=A,B.
\label{eq:lj}
\end{equation}
Here, the LJ energy parameters are set as 
$\epsilon_{AA}=\epsilon_{BB}=2\epsilon_{AB}=\epsilon$. The details of 
equilibrium phase behavior for this potential are well studied 
\cite{dhb03,dhbfs06,dfshb06}. 
If we express all lengths 
in terms of the LJ diameter $\sigma$, masses in units of $m$ ($m_A=m_B=m$), and energies 
in terms of $\epsilon$, the natural time unit is
\begin{equation}
 t_0=\sqrt{\frac{m\sigma^2}{48\epsilon}}.
 \label{eq:mdtime}
\end{equation}
Setting $\sigma=1$, $m=1$, and $\epsilon=1$ gives $t_0=1/\sqrt{48}$. The potential in Eq.~(\ref{eq:lj})
is cut-off at $r_c=2.5\,\sigma$ to enhance computational speed. To remove the discontinuities 
in the potential and force at $r=r_c$, we invoke the {\it shifted potential} and 
{\it shifted-force potential} corrections to the potential in Eq.~(\ref{eq:lj}) \cite{at87}.

For the potential between the walls and the fluid particles, we consider 
an integrated LJ potential ($\alpha=A,B$):
\begin{equation}
u_w(z)=\frac{2\pi n\sigma^3}{3} \left[\frac{2\epsilon_r}{15}{\left(\frac{\sigma}{z^\prime}\right)}^9
-\delta_\alpha\epsilon_a{\left(\frac{\sigma}{z^\prime}\right)}^3\right].
\label{eq:intlj}
\end{equation}
Here, $n$ is the reference density of the bulk fluid, and 
$\epsilon_r$ and $\epsilon_a$ are the energy scales for the repulsive 
and attractive parts of the interaction. We set $\delta_A=1$ and 
$\delta_B=0$ for the wall at $z=0$. Thus, $A$ particles are attracted at large 
distances and repelled at short distances, whereas $B$ particles feel only repulsion. 
For the wall at $z=D$, we choose $\delta_A=0$ and $\delta_B=0$, so that there is only 
a repulsion for both $A$ and $B$ particles. Furthermore, we have
$z^\prime=z+\sigma/2$ for the wall at $z=0$, and $z^\prime=D+\sigma/2-z$ for
the wall at $z=D$. We notice that this simplified potential incorporates the effect of a semi-infinite geometry (the generalization to any other geometry is straightforward). However, it does not take into account the surface structure in the $xy$-plane.

The fluid has $N=N_A+N_B$ particles, and the fluid density is $n=N/(L_w^2 D)=1$. In our simulations, we chose $L_w=48$ and $D=48$ ($N=110592$ particles). For the range of times studied here ($t \leq 2800$), test runs with other values of $L_w$ showed that $L_w = 48$ is large enough to ensure that the laterally inhomogeneous domains that form during coarsening are not affected by finite-size effects. The statistical quantities presented here were obtained as averages over $50$ independent runs. We performed simulations 
on the fluid for the surface potential [Eq.~(\ref{eq:intlj})] with 
$\epsilon_a=0.1,0.6$, while $\epsilon_r=0.5$. We find that 
$\epsilon_a=0.1$ corresponds to a PW morphology, while $\epsilon_a=0.6$ yields 
a CW morphology \cite{db10}. The quench temperature is $T=1.0\simeq 0.7T_c$ (bulk $T_c\simeq 1.423$) 
\cite{dhbfs06,dfshb06}, and is maintained by the Nos\'{e}-Hoover thermostat which preserves hydrodynamics \cite{adp10,bc96}. The homogeneous initial state of the fluid mixture is prepared from a short run at high $T$ ($\gg T_c$), with periodic boundary conditions imposed in all directions. Finally, Newton's
equations of motion are integrated numerically using the Verlet velocity algorithm 
\cite{bc96}, with a time-step $\Delta t=0.07$ in LJ units.

We undertook extensive MD simulations to study the time-dependent 
morphology which arises during surface-directed phase separation. 
We characterized the morphology via layer-wise correlation functions, 
structure factors, and length-scales. We also computed laterally-averaged order parameter profiles 
and their various properties, e.g., surface value of the order parameter, zero-crossings, etc. Before presenting these quantities, it is useful to summarize theoretical results in this context. 

\section{Theoretical Background}
\label{sec:theory}

\subsection{Kinetics of Phase Separation in the Bulk}
\label{subsec:bulk}

The coarsening domains have a characteristic length scale 
$L(t)$, which grows with time. For pure and isotropic systems, 
$L(t)\sim t^\theta$, where the growth exponent $\theta$ depends
on the conservation laws, the nature of defects which drive the 
evolution, and the relevance of hydrodynamic flow fields. 

First, we discuss the domain growth laws which arise in bulk phase-separating systems
\cite{ls61,es79,hf85,bs74,kb77,huse86}.
For diffusive dynamics, the order parameter satisfies the {\it Cahn-Hilliard} (CH)
equation. In dimensionless variables, this has the form
\cite{pw09}
\begin{equation}
 \frac{\partial}{\partial t}\psi(\vec{r},t)=\nabla^2
\left(-\psi+\psi^3-\frac{1}{2}\nabla^2\psi\right) ,
\label{eq:ch}
\end{equation}
where the order parameter $\psi(\vec{r},t)$ is proportional to the $AB$ density
difference at space-point $\vec{r}$ and time $t$. Lifshitz and Slyozov (LS) 
\cite{ls61} 
considered the diffusion-driven growth of a droplet of the minority phase in
a supersaturated background of the majority phase. The LS 
mechanism leads to the growth law $L(t)\sim t^{1/3}$ in $d\geq 2$. Huse \cite{huse86} 
argued that this law is also valid for spinodal decomposition in
mixtures with approximately equal fractions of the two components. Typically, 
for a domain of size $L$, the chemical potential on its surface is 
$\mu\sim\sigma/L$, where $\sigma$ is the surface tension. Then the current is 
$D|\vec{\nabla}\mu|\sim D\sigma/L^2$, where $D$ is the diffusion 
constant. Therefore, the domain size grows as $dL/dt\sim D\sigma/L^2$, or 
$L(t)\sim(D\sigma t)^{1/3}$.

Next, we consider the segregation of binary fluids, where 
the hydrodynamic flow field provides an additional mechanism for 
transport of material \cite{pw09,bray94,bf01,onuki02}. 
Hydrodynamic effects can be incorporated in the CH model 
by including a velocity field which satisfies the Navier-Stokes equation -- the resultant coupled equations are termed as {\it Model H} \cite{hh77}. 
The growth dynamics is diffusion-limited at early times, as in the case 
of binary alloys. However, one finds a crossover to a hydrodynamic growth 
regime, where convection assists in the rapid transportation of material 
along the domain boundaries \cite{es79,hf85}. The growth laws for different regimes are summarized as follows \cite{pw09}:
\begin{eqnarray}
L(t) &\sim& (D\sigma t)^{1/3} , \quad L \ll (D \eta)^{1/2} ,
\quad \mbox{({diffusive regime})} \nonumber \\
&\sim& \frac{\sigma t}{\eta} , \quad (D \eta)^{1/2} \ll
L \ll \frac{\eta^2}{\rho \sigma} , \quad \mbox{({viscous hydrodynamic regime})}
\nonumber \\
&\sim& \left( \frac{\sigma t^2}{\rho} \right)^{1/3} , \quad
\frac{\eta^2}{\rho \sigma} \ll L, \quad \mbox{({inertial hydrodynamic regime})} .
\label{eq:growth} 
\end{eqnarray}
In Eq.~(\ref{eq:growth}), $\eta$ and $\rho$ denote the viscosity and density of the 
fluid, respectively.

\subsection{Kinetics of Phase Separation at Wetting Surfaces}
\label{subsec:surface}

Next, we briefly discuss phase-separation kinetics at wetting surfaces \cite{puri05,bpdh10}. For the diffusive case,
the order parameter satisfies the CH equation in the bulk:
\begin{equation}
 \frac{\partial}{\partial t}\psi(\vec{\rho},z,t)=\nabla^2
\left[-\psi+\psi^3-\frac{1}{2}\nabla^2\psi+V(z)\right], \quad z>0.
\label{eq:bulk}
\end{equation}
In Eq.~(\ref{eq:bulk}), we have designated $\vec{r}\equiv(\vec{\rho},z)$, where $\vec{\rho}$ and $z$ denote 
coordinates parallel and perpendicular to the surface (located at $z=0$), 
respectively. The surface potential $V(z)$ is chosen such that the surface 
preferentially attracts $A$. 

Equation~(\ref{eq:bulk}) must be supplemented by two boundary 
conditions at $z=0$
\cite{pb92,puri05},
as it is a fourth-order partial differential 
equation. Now, since the surface value of the order parameter is not conserved, we assume a nonconserved relaxational kinetics for this quantity:
\begin{equation}
 \frac{\partial}{\partial t}\psi(\vec{\rho},0,t)=h_1+g\psi(\vec{\rho},0,t)
+\gamma \frac{\partial}{\partial z}\psi(\vec{\rho},z,t)\bigg|_{z=0}
+\tilde{\gamma}\nabla_\|^2\psi(\vec{\rho},0,t) .
\label{eq:bc1}
\end{equation}
In Eq.~(\ref{eq:bc1}), $h_1=-V(0)$, and $g,\gamma,\tilde{\gamma}$ are phenomenological 
parameters; and $\nabla_\|^2$ denotes the in-plane Laplacian.
Next, we implement a zero-current boundary condition at 
the surface, which enforces the conservation of the order parameter:
\begin{eqnarray}
 0=\dfrac{\partial}{\partial z}\Bigg[ 
 -\psi+\psi^3 
 -\frac{1}{2}\nabla^2\psi+V(z)\Bigg]\Bigg|_{z=0}. 
 \label{eq:bc2}
\end{eqnarray}
Equations~(\ref{eq:bulk})-(\ref{eq:bc2}) describe the kinetics of SDSD 
with diffusive dynamics. This is appropriate for phase separation in solid 
mixtures, or the early stages of segregation in polymer blends. 
However, most experiments involve fluid mixtures, where hydrodynamics 
plays an important role in the intermediate and late stages 
of phase separation. At a phenomenological level, hydrodynamic effects
can be incorporated via the Navier-Stokes equation for the velocity field  
\cite{hh77}. This must be supplemented by appropriate boundary 
conditions at the surfaces \cite{tanaka01}. Alternatively, we can consider 
molecular models of fluid mixtures at a surface, in which the fluid velocity 
field is naturally included. We adopt the latter strategy 
in this paper and study SDSD in fluid mixtures via MD simulations.

Let us briefly discuss the growth laws which arise in SDSD. 
At early times, the wetting-layer growth is driven by the diffusion of $A$ particles 
from bulk domains of size $L\sim{(\sigma t)}^{1/3}$ (with $\mu \sim \sigma/L$)
to the flat surface layer of size $\simeq\infty$ 
(with $\mu\simeq 0$). Therefore, neglecting the contribution due to the surface potential 
at very early times 
\cite{pb01}, 
we obtain
\begin{equation}
 \frac{dR_1}{dt} \sim \frac{\sigma}{Lh} \sim \frac{\sigma}{LR_1}.
\label{eq:R1}
\end{equation}
In Eq.~(\ref{eq:R1}), $h\sim R_1$ is the thickness of the depletion layer. The LS growth 
law for the wetting-layer thickness [$R_1\sim{(\sigma t)}^{1/3}$] can be readily obtained 
from Eq.~(\ref{eq:R1}). At later times, $R_1$ shows a rapid growth due to the 
establishment of contact between the bulk tubes and the wetting layer. Then, the wetting component
is pumped hydrodynamically to the surface. The subsequent growth dynamics is similar to 
that in segregation of fluids. We expect $R_1(t)\sim t$ in the viscous hydrodynamic 
regime, followed by a crossover to $R_1(t)\sim t^{2/3}$  in the inertial hydrodynamic regime. 

\section{Detailed Numerical Results}
\label{sec:results}

\begin{figure}[!htbp]
\centering
\begin{tabular}{c}
{\includegraphics*[width=0.75\textwidth]{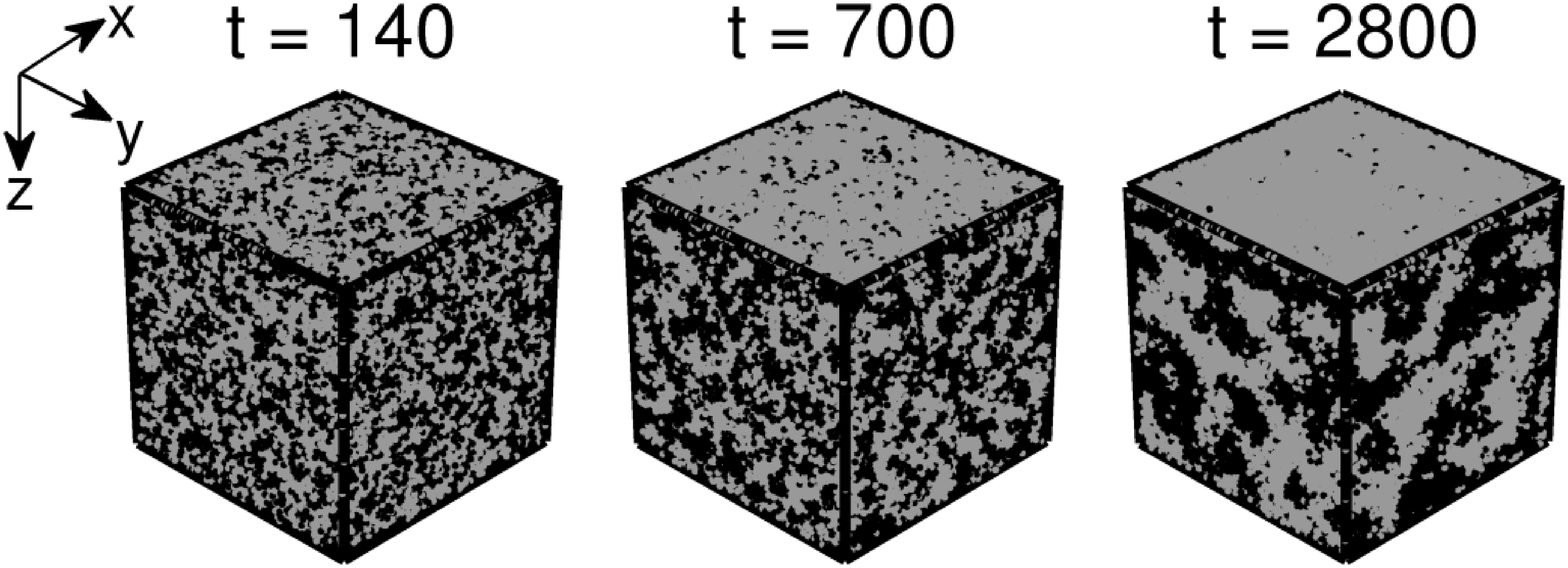}} \\ 
{\includegraphics*[width=0.75\textwidth]{./fig1b.eps}} \\ 
\end{tabular}
\caption{Evolution snapshots (upper frames) for surface-directed spinodal decomposition 
(SDSD) in a binary ($AB$) Lennard-Jones mixture, which is confined in a box of size 
$L_w^2\times D$, with $L_w=48$, $D=48$. 
An impenetrable surface (located at $z=0$) attracts the $A$-particles (marked gray). 
The surface field strength is given by $\epsilon_a = 0.6$ and $\epsilon_r=0.5$
in Eq.~(\ref{eq:intlj}), which corresponds to a
completely wet (CW) morphology in equilibrium. The temperature is 
$T=1.0\simeq 0.7\,T_c\, (\mbox{bulk}~T_c=1.423)$. The other simulation details are 
provided in the text. The $B$-particles are marked black. The lower frames
show the $yz$-cross-sections of the upper frames at $x=0$.}
\label{fig:fig1}
\end{figure}

In this section, we present results from our MD simulations. The details of these have been described in Sec.~\ref{sec:md}. First, we focus on domain morphologies and laterally-averaged profiles for the CW case. In Fig.~\ref{fig:fig1}, we show evolution snapshots and their $yz$-cross-sections for SDSD in a binary ($AB$) fluid mixture at different times. The surface field strengths are $\epsilon_r=0.5,\epsilon_a=0.6$ in 
Eq.~(\ref{eq:intlj}), which correspond to a CW morphology in equilibrium. 
An $A$-rich layer develops at the surface ($z=0$), resulting 
in SDSD waves which propagate into the bulk. Consequently, the surface 
exhibits a multi-layered morphology, i.e., wetting layer followed by 
depletion layer, etc. The snapshots (and their cross-sections in the lower frames) clearly show that only $A$-particles are at the surface, as expected for a CW morphology.

\begin{figure}[!htbp]
\centering
\includegraphics*[width=0.7271\textwidth]{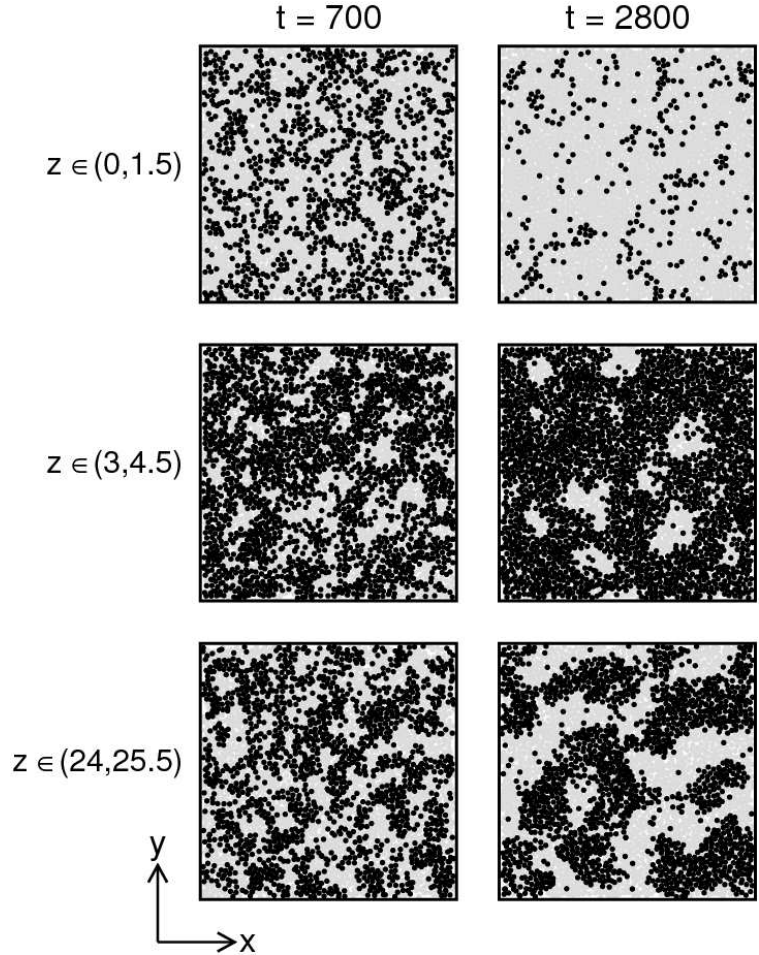}
\caption{Cross-section slices of size $L_w^2\times 1.5\sigma$ for 
the evolution shown in Fig.~\ref{fig:fig1} at $t=700,2800$ MD units. The slices 
show all $A$ atoms (marked in gray) and all $B$ atoms (marked in black) lying 
in the interval $z \in (0,1.5)$ (top frames), $z \in (3,4.5)$ (middle frames), $z \in (24,25.5)$ (bottom frames).}
\label{fig:fig2}
\end{figure}

\begin{figure}[!htbp]
\centering
\includegraphics*[width=0.7\textwidth]{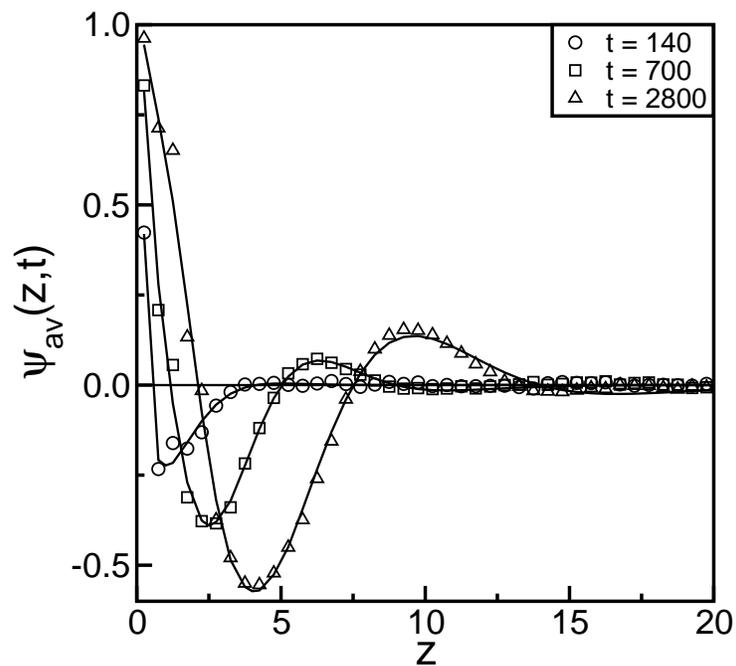}
\caption{Laterally-averaged order parameter profiles for the evolution 
shown in Fig.~\ref{fig:fig1} at $t=140, 700, 2800$ MD units. 
The continuous lines through the data points are guides to the eye.}
\label{fig:fig3}
\end{figure}

In Fig.~\ref{fig:fig2}, we show cross-sections in the $xy$-plane for the 
evolution snapshots in Fig.~\ref{fig:fig1}. The surface layer (shown in
the top frames at $t=700,2800$) has almost no $B$-particles. In the middle frames, we notice that 
there is a surplus of $B$ atoms due to the migration of $A$ to the surface. 
(This is confirmed by the laterally-averaged profiles, shown in Fig.~\ref{fig:fig3}.)
The bottom frames show the usual segregation morphologies in the bulk -- they
correspond to the region $z\in[24,25.5]$, which is unaffected by the SDSD
waves at these simulation times (see Fig.~\ref{fig:fig3}).

Depth-profiling techniques in experiments do not have much lateral resolution, and yield 
only laterally-averaged order parameter profiles $\psi_{\textrm{av}}(z,t)$ vs. $z$ 
\cite{gk03}. The numerical counterpart of these profiles is obtained by 
averaging $\psi(\vec{\rho},z,t)$ along the $x,y$ directions, and then further 
averaging over 50 independent runs. The order parameter is defined from the
local densities $n_A, n_B$ as
\begin{equation}
 \psi(\vec{r},t)=\frac{n_A - n_B}{n_A + n_B}.
 \label{eq:opdef}
\end{equation}
In Fig.~\ref{fig:fig3}, we show the depth profiles 
for the evolution depicted in Fig.~\ref{fig:fig1}. Figure~\ref{fig:fig3} 
clarifies the nature of the multi-layered morphology seen in SDSD. In the bulk, the 
SDSD wave-vectors are randomly oriented, which results in $\psi_{\textrm{av}}(z,t) 
\simeq 0$ due to the averaging procedure. However, the averaged profiles show a 
systematic oscillatory behavior at the surface. 

\begin{figure}[!htbp]
\centering
\includegraphics*[width=0.75\textwidth]{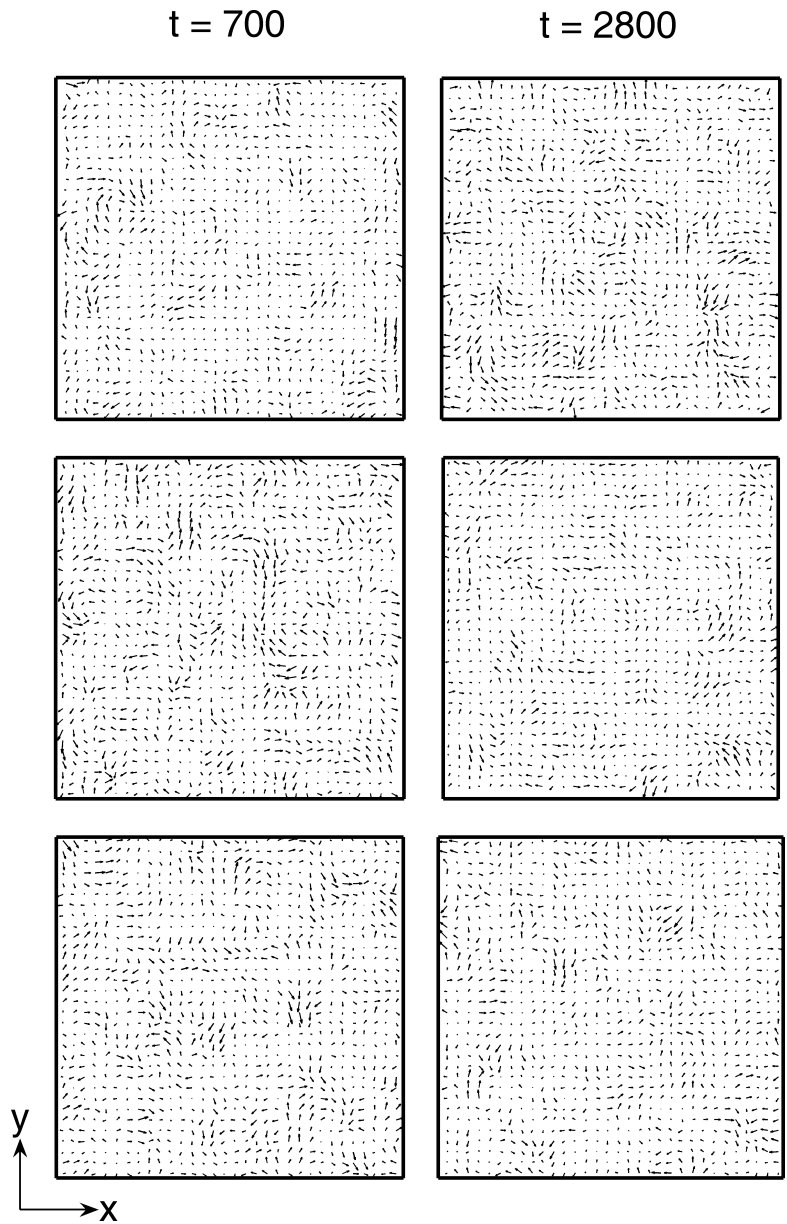}
\caption{Analogous to Fig.~\ref{fig:fig2}, but for the velocity field
($v_x,v_y$) in the $xy$-plane. The velocities are coarse-grained as described
in the text.}
\label{fig:fig4}
\end{figure}

Let us next examine the velocity field at the surface and in the bulk. In Fig.~\ref{fig:fig4},
we show the ($v_x,v_y$)-field in the $xy$-planes used in Fig.~\ref{fig:fig2}.
The snapshots shown in Fig.~\ref{fig:fig4} are obtained by coarse-graining the velocities 
in overlapping boxes of size $(4.5\sigma)^3$. These boxes are centred on cubes of size
$(1.5\sigma)^3$, and we show the ($v_x,v_y$)-field for these cubes. We make the following observations concerning Fig.~\ref{fig:fig4}: \\
1) The velocity field is characterized by vortices and anti-vortices, but these
do not show much coarsening with time -- compare the snapshots at time
$t=700,2800$ for different values of $z$. This has also been observed in MD
studies of bulk spinodal decomposition by Ahmad et al. \cite{adp10}. \\
2) There are no significant morphological differences between the velocity fields
at the surface (top frames of Fig.~\ref{fig:fig4}) and in the bulk (bottom frames of
Fig.~\ref{fig:fig4}). This is confirmed by comparing the corresponding correlation
functions -- for brevity, we do not present these here.

\begin{figure}[!htbp]
\centering
\includegraphics*[width=0.75\textwidth]{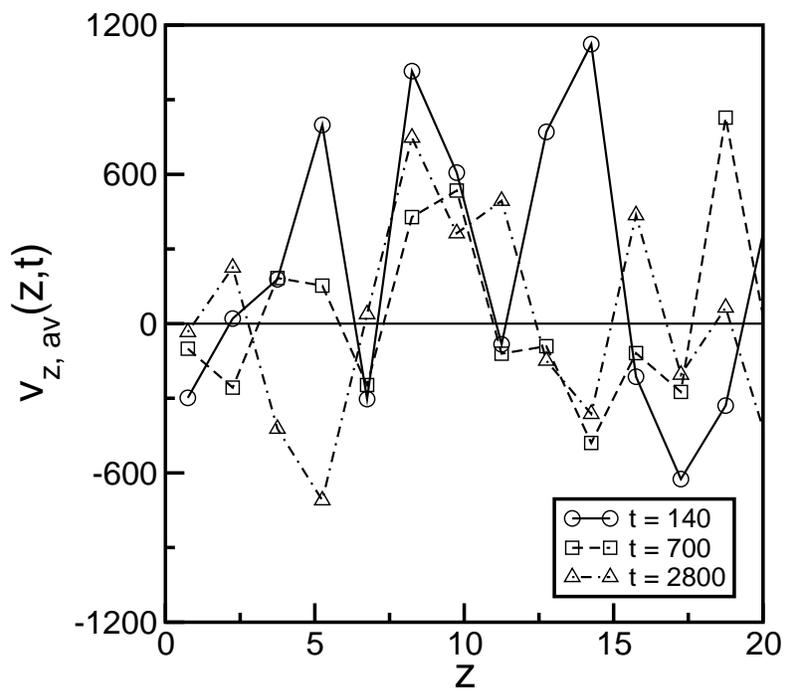}
\caption{Laterally-averaged $v_z$-profiles for the evolution 
shown in Fig.~\ref{fig:fig1} at $t=140, 700, 2800$ MD units.}
\label{fig:fig5}
\end{figure}

It is relevant to ask whether the depth profiles of the velocity field show any
systematic behavior (as in Fig.~\ref{fig:fig3}). In Fig.~\ref{fig:fig5}, we plot
$v_{z,\textrm{av}}(z,t)$ vs. $z$ for $t=140,700,2800$. The procedure for
calculating the laterally-averaged velocity field is as follows: in each layer
of thickness $1.5 \sigma$ (along the $z$ direction), we sum up the $z$-component
of the velocities for all particles. Clearly, the depth profiles of the velocity
field do not show any major systematic features.

\begin{figure}[!htbp]
\centering
\begin{tabular}{c}
{\includegraphics*[width=0.75\textwidth]{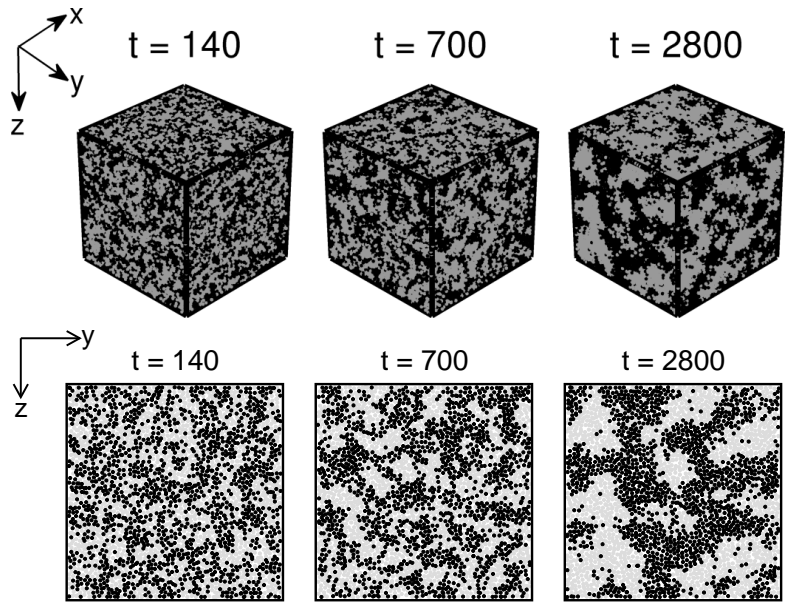}} \\ 
{\includegraphics*[width=0.75\textwidth]{./fig6b.eps}} \\ 
\end{tabular}
\caption{Analogous to Fig.~\ref{fig:fig1}, but for the case with $\epsilon_a = 0.1$ and $\epsilon_r = 0.5$. These parameters correspond to a partially wet (PW) morphology in equilibrium.}
\label{fig:fig6}
\end{figure}

Next, we turn our attention on the morphologies and profiles for the PW case. 
The evolution snapshots and their $yz$-cross-sections for the PW morphology are 
shown in Fig.~\ref{fig:fig6}. In this case, we set $\epsilon_r=0.5,\epsilon_a=0.1$ 
in Eq.~(\ref{eq:intlj}). As in the CW case, we again observe usual phase-separation 
morphologies in the bulk. However, in this case, both $A$ and $B$ particles are
present at the surface. 

\begin{figure}[!htbp]
\centering
\includegraphics*[width=0.7271\textwidth]{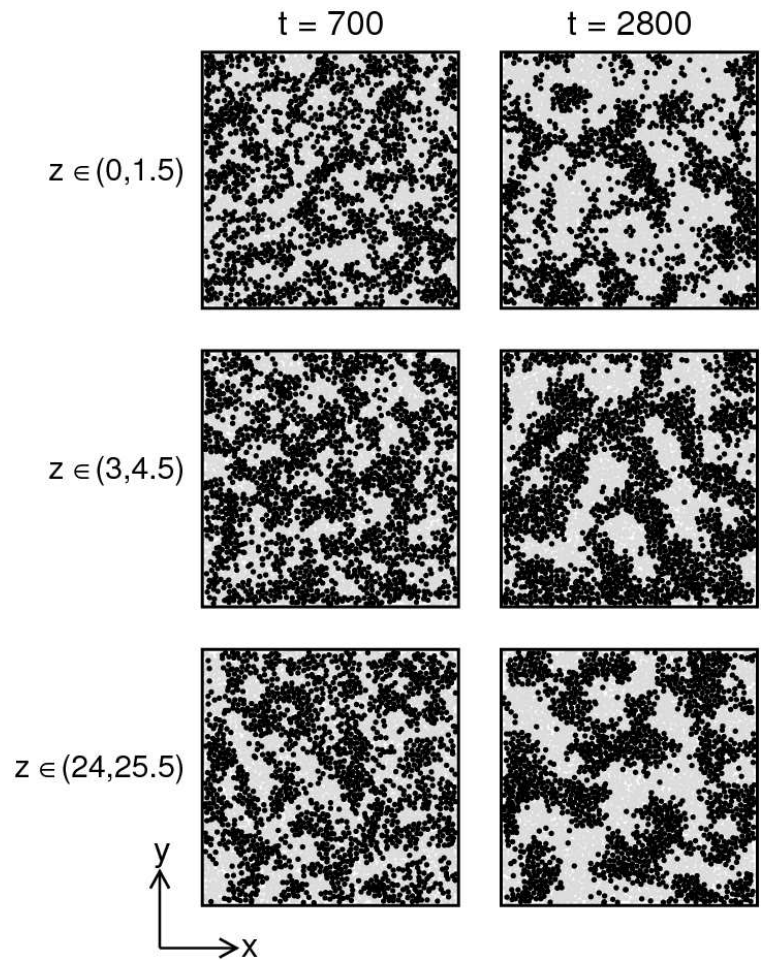}
\caption{Analogous to Fig.~\ref{fig:fig2}, but for the evolution shown 
in Fig.~\ref{fig:fig6}.}
\label{fig:fig7}
\end{figure}

\begin{figure}[!htbp]
\centering
\includegraphics*[width=0.7\textwidth]{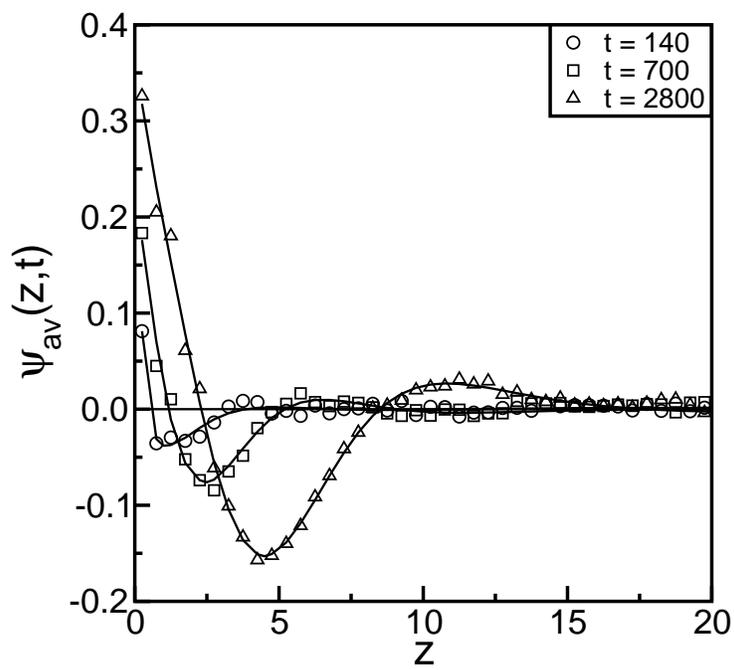}
\caption{Analogous to Fig.~\ref{fig:fig3}, but corresponding to the evolution shown 
in Fig.~\ref{fig:fig6}.}
\label{fig:fig8}
\end{figure}

Figure~\ref{fig:fig7} shows the cross-sections in the $xy$-plane, corresponding to 
the evolution in Fig.~\ref{fig:fig6}. At early times ($t=700$, top frame),
approximately equal numbers of $A$ and $B$ particles are present at the surface.
However, there is a surplus of $A$ atoms at late times ($t=2800$, top frame),
as expected in the PW morphology. In the middle frames, we see more $B$ particles,
as $A$ atoms have migrated to the surface. The laterally-averaged profiles in
Fig.~\ref{fig:fig8} show that $z \in [3,4.5]$ (corresponding to the middle frames
in Fig.~\ref{fig:fig7}) lies in the depletion layer for both $t=700,2800$.
The bottom frames in Fig.~\ref{fig:fig7} show the segregation kinetics in the bulk.

We plot $\psi_{\textrm{av}}(z,t)$ vs. $z$ in Fig.~\ref{fig:fig8}, corresponding to 
the PW evolution in Fig.~\ref{fig:fig6}. A behavior similar to the CW morphology
(cf. Fig.~\ref{fig:fig3}) is seen in this case too. However, notice that the degree
of surface enrichment (and depletion adjacent to the surface) is much less in
Fig.~\ref{fig:fig8}.

We have also studied the morphology of the velocity field in the PW case. The
features are analogous to those in Figs.~\ref{fig:fig4} and \ref{fig:fig5} for the
CW case, and we do not show these results here.

\begin{figure}[!htbp]
\centering
\includegraphics*[width=0.75\textwidth]{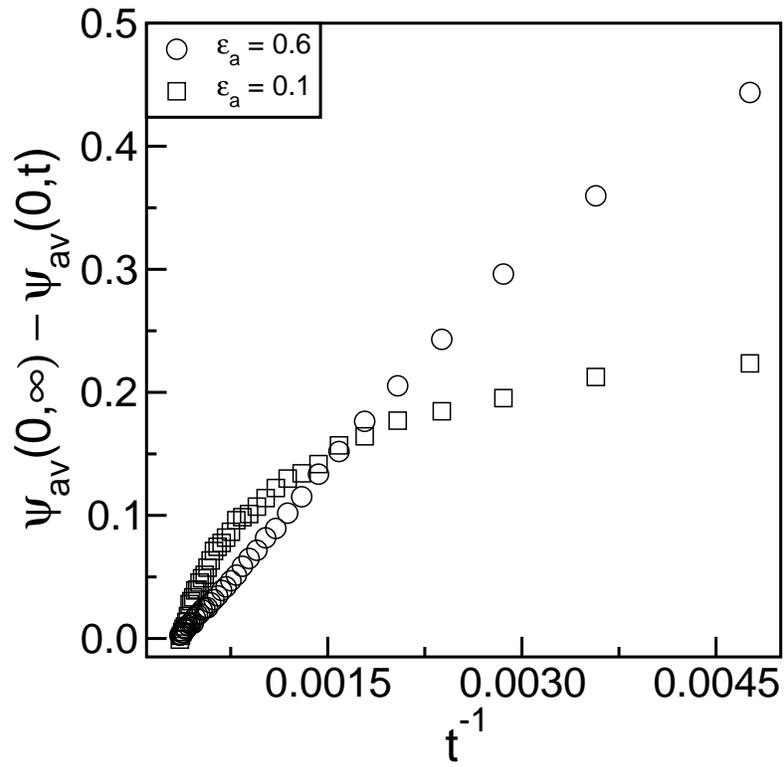}
\caption{Time-dependence of the surface value of the order parameter 
for the CW and PW profiles in Figs.~\ref{fig:fig3} and \ref{fig:fig8},
respectively. We plot 
$\psi_{\textrm{av}}(0,\infty)-\psi_{\textrm{av}}(0,t)$ vs. $t^{-1}$.}
\label{fig:fig9}
\end{figure}

Next, let us examine some quantitative properties of the depth profiles in Figs.~\ref{fig:fig3} and \ref{fig:fig8}.
Figure~{\ref{fig:fig9}} shows the time-dependence 
of the surface value of the order parameter for the CW and PW cases. We plot 
$\psi_{\textrm{av}}(0,\infty)-\psi_{\textrm{av}}(0,t)$ vs. $t^{-1}$, demonstrating that $\psi_{\textrm{av}}(0,t)$ 
saturates linearly to its asymptotic value $\psi_{\textrm{av}}(0,\infty)$ for the CW case (with $\epsilon_a = 0.6$):
\beq
\psi_{\textrm{av}}(0,t) \simeq \psi_{\textrm{av}}(0,\infty) - \frac{A}{t} + \ldots ,
\eeq
where $A$ is a constant. Notice that the asymptotic value $\psi_{\textrm{av}}(0,\infty)$ is estimated by extrapolation of the data for $\psi_{\textrm{av}}(0,t)$ vs. $t$. The corresponding behavior for the PW case (with $\epsilon_a = 0.1$) is not so clear. However, our results suggest that the PW case also saturates linearly at long times.

\begin{figure}[!htbp]
\centering
\includegraphics*[width=0.75\textwidth]{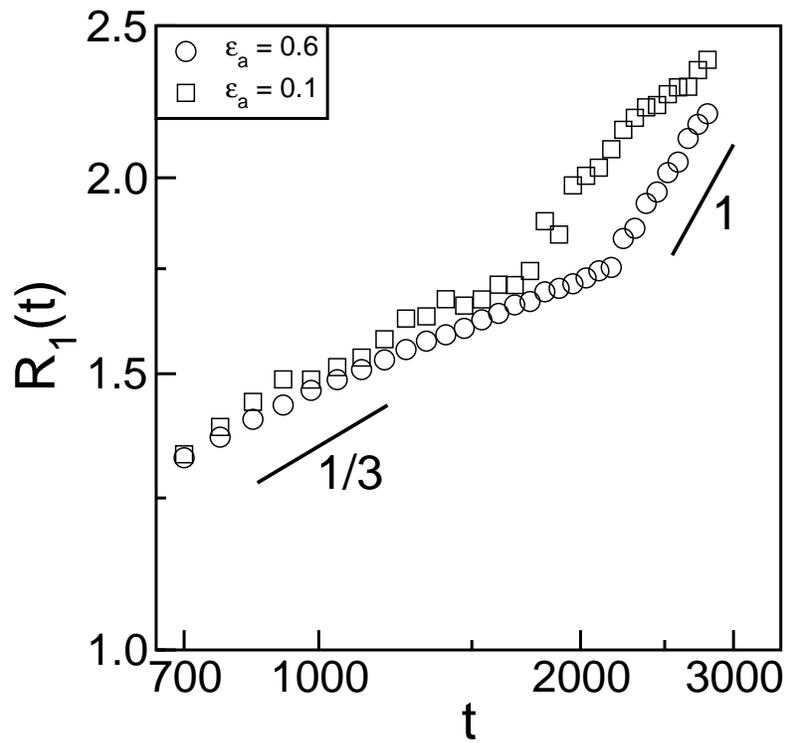}
\caption{Time-dependence of the wetting-layer thickness $R_1(t)$ of the CW and PW 
profiles on a log-log scale. The straight lines have slopes $1/3$ and $1$, corresponding to 
the diffusive regime and the viscous hydrodynamic regime, respectively.}
\label{fig:fig10}
\end{figure}

The evolution of the SDSD profiles in Figs.~\ref{fig:fig3} and \ref{fig:fig8} is
characterized by the zero-crossings of $\psi_{\textrm{av}}(z,t)$. 
The quantity $R_1(t)$ denotes the first zero, and measures the wetting-layer
thickness. Figure~\ref{fig:fig10} plots $R_1(t)$ vs. $t$ for the CW and PW
cases shown in Figs.~\ref{fig:fig3} and \ref{fig:fig8}.
This plot shows a power-law behavior for the growth dynamics, 
$R_1(t)\sim t^\theta$, but there is a distinct crossover in the growth exponent. 
For $t\leq t_c \simeq 2000$, we have $\theta\simeq 1/3$, in conformity with the LS mechanism 
for diffusive growth. However, for $t\geq t_c$, we observe a much more rapid growth 
with $\theta\simeq 1$, corresponding to the viscous hydrodynamic regime.
We make the following observations regarding Fig.~\ref{fig:fig10}: \\
1) The crossover time is consistent with the observation
of a $1/3 \to 1$ crossover (at $t_c \simeq 2000$) in bulk MD simulations by
Ahmad et al. \cite{adp10}. These authors used a similar model, but without surface interactions. \\
2) The crossover in the CW case is much sharper than in the PW case. In the
CW case, bulk tubes establish contact with a flat wetting layer, and rapidly drain
into it. In the PW case, the surface morphology consists of semi-droplets, and the
pressure differences from the bulk tubes are less marked. \\
3) We can go up to $t\simeq 3000$ for these system sizes ($L_w=48, D=48$). Beyond 
this time, the system encounters finite-size effects due to the lateral domain size 
becoming an appreciable fraction of the system size $L_w$. Presently, our computational 
constraints do not allow us to access the inertial hydrodynamic regime (with $\theta = 2/3$) via MD simulations 
\cite{adp10}. However, our results for the wetting-layer dynamics show the viscous hydrodynamic regime, though in a limited time-window.

Before concluding this section, we discuss some other quantitative features of the domain morphologies. We present results for the CW case only -- the PW results are analogous. First, we focus on the layer-wise correlation function, which characterizes the domain morphology. This is defined as follows 
\cite{pb94}
\begin{equation}
 C_\parallel (\vec{\rho},z,t) = L_w^{-2}\int d\vec{\sigma}\left[\langle\psi(\vec{\sigma},z,t)
 \psi(\vec{\sigma}+\vec{\rho},z,t)\rangle-\langle\psi(\vec{\sigma},z,t)\rangle
 \langle\psi(\vec{\sigma}+\vec{\rho},z,t)\rangle\right],
 \label{eq:lcf}
\end{equation}
where the angular brackets denote statistical averaging over independent runs. We 
denote $C_\parallel (\vec{\rho},z,t)$ as $C(\vec{\rho},t)$ in the following discussion 
for convenience. Since the system is isotropic in the $x,y$ directions, $C$ is 
independent of the direction of $\vec{\rho}$. We can define the $z$-dependent lateral 
length scale $L_\parallel(z,t)\equiv L(z,t)$ from the half-decay of $C(\rho,t)$ 
\cite{pb94}:
\begin{equation}
 C(\rho=L,t)=\frac{1}{2}C(0,t).
 \label{eq:llen}
\end{equation}
To obtain the correlation function, etc., a coarse-graining procedure \cite{dp02} 
is employed, which is the numerical counterpart of the renormalization group (RG) technique. 
We divide our system into small boxes of size 
$\sigma^2\times 1.5\sigma$. We count the total number of $A$ and $B$ particles in 
each box and its nearest neighbors. If there are more particles of $A$ than $B$ in the box and its neighbors, 
we assign a ``spin'' value $S=+1$ to that box. On the other hand, the box is given a spin value $S=-1$ when 
there are more $B$ particles than $A$. Furthermore, we assign $+1$ or $-1$ to a box randomly, 
when equal numbers of $A$ and $B$ particles are present. 

\begin{figure}[!htbp]
\centering
\includegraphics*[width=0.75\textwidth]{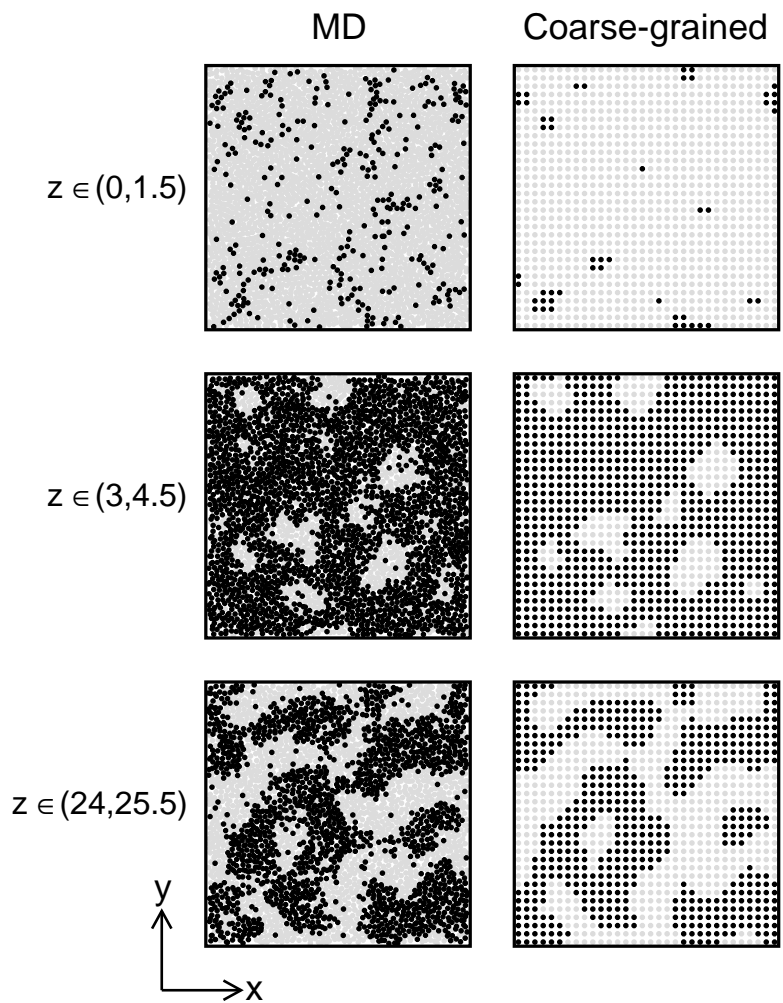}
\caption{Cross-sections of the SDSD snapshots (frames on left) at $t=2800$ shown in 
Fig.~\ref{fig:fig2}. The cross-sections show all $A$ atoms (marked gray) and 
all $B$ atoms (marked black) lying in the interval $z \in (0,1.5)$ (top frame), 
$z \in (3,4.5)$ (middle frame), $z \in (24,25.5)$ (bottom frame). The frames on the
right show coarse-grained versions of the MD snapshots. The coarse-graining
procedure is described in the text.}
\label{fig:fig11}
\end{figure}

The results of this coarse-graining procedure are shown in Fig.~\ref{fig:fig11}. In the frames on the left, we reproduce the $xy$-cross-sections of the SDSD snapshots at $t=2800$ in Fig.~\ref{fig:fig2}. The frames on the right show the corresponding coarse-grained pictures. Figure~\ref{fig:fig11} clearly demonstrates the elimination of fluctuations in our coarse-grained snapshots, while preserving the important morphological features. 

\begin{figure}[!htbp]
\centering
\includegraphics*[width=0.47\textwidth]{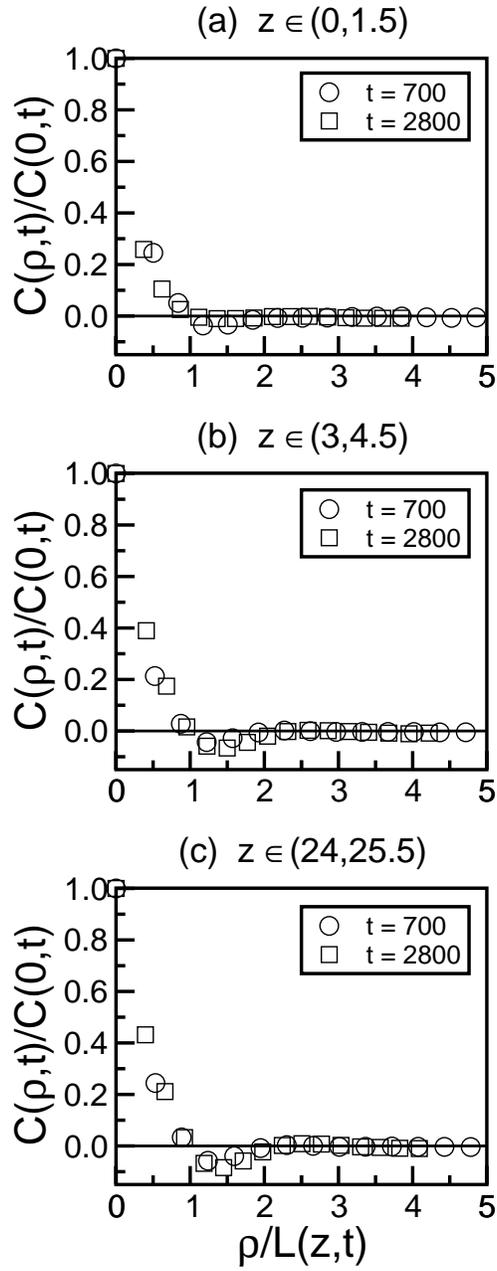}
\caption{Scaling plot of the layer-wise correlation functions for the CW evolution depicted in 
Fig.~\ref{fig:fig1}. We plot $C(\rho,t)/C(0,t)$ vs. $\rho/L(z,t)$ for $t=700,2800$ with
(a) $z \in (0,1.5)$; (b) $z \in (3,4.5)$; (c) $z \in (24,25.5)$.}
\label{fig:fig12}
\end{figure}

In Fig.~\ref{fig:fig12}, we plot the normalized correlation function $C(\rho,t)/C(0,t)$ 
(computed from the coarse-grained spin variable) vs. $\rho/L(z,t)$ for three 
different layers, as indicated in the figure. The surface layer [$z\in (0,1.5)$]
has few inhomogeneities, and shows a corresponding lack of structure in the 
correlation function. [Notice that a state with $S_i=+1$ $\forall i$ has
$C(\rho)=0$ from our definition in Eq.~(\ref{eq:lcf}).] The layer at
$z\in (3,4.5)$ lies in the depletion region for $t=700,2800$, as is seen
from the laterally-averaged profiles in Fig.~\ref{fig:fig3}. The corresponding
correlation functions (middle frame of Fig.~\ref{fig:fig12}) show scaling
behavior. The bottom frame in Fig.~\ref{fig:fig12} corresponds to a bi-continuous
bulk morphology -- see bottom frames of Fig.~\ref{fig:fig11}.

\begin{figure}[!htbp]
\centering
\includegraphics*[width=0.75\textwidth]{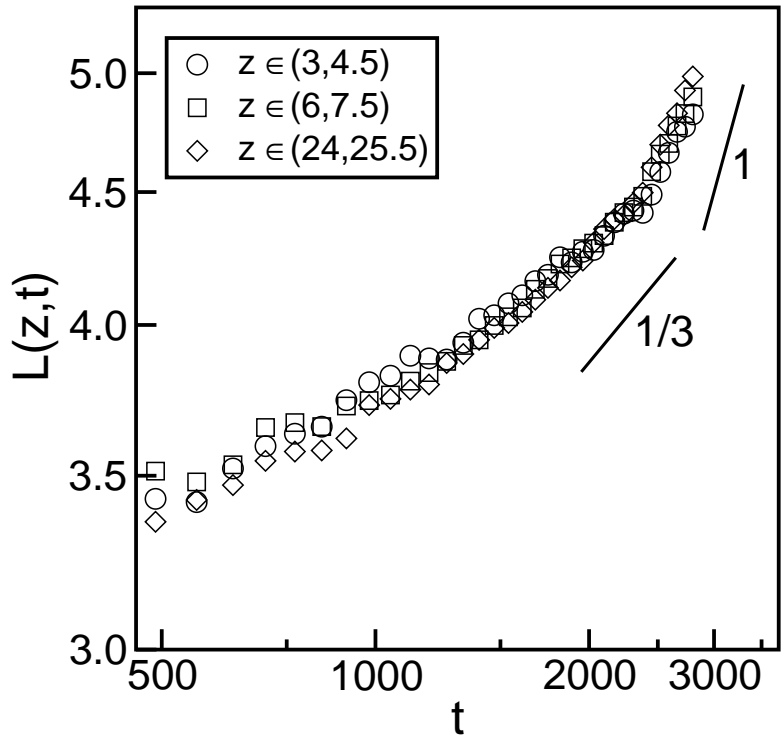}
\caption{Time-dependence of the layer-wise length scale for the evolution 
depicted in Fig.~\ref{fig:fig1}. We plot $L(z,t)$ vs. $t$ on a log-log scale 
for various values of $z$. The solid lines have slopes $1/3$ (diffusive regime) and $1$ 
(viscous hydrodynamic regime).}
\label{fig:fig13}
\end{figure}

Finally, we focus on the time-dependence of the lateral domain size $L(z,t)$. In 
Fig.~\ref{fig:fig13}, we plot $L(z,t)$ vs. $t$ for three different layers, excluding
the surface layer. (As is evident from the top frames of Fig.~\ref{fig:fig11},
there is no characteristic ``domain scale'' associated with the surface layer.) We find 
that $L(z,t)$ grows as a power-law with time ($L\sim t^\theta$), but there is a 
crossover in the growth exponent. The early-time dynamics ($t\leq t_c$) is 
consistent with the expected diffusive LS growth law with $\theta \simeq 1/3$  
\cite{pw09,bray94,bf01,onuki02}. However, there is a much more rapid growth at late times 
($t\geq t_c$) with $\theta \simeq 1$. Notice that the crossover time ($t_c\simeq 2000$)
is consistent with the crossover time for the growth dynamics of the wetting layer.

\section{Summary and Discussion}
\label{sec:sum}

Let us conclude this paper with a brief summary and discussion of our results. We have 
studied {\it surface-directed spinodal decomposition} (SDSD) in an unstable 
homogeneous binary ($AB$) mixture at a wetting surface ($S$). Depending on the relative 
values of the surface tensions between $A,B$ and $S$, the equilibrium morphology can be either 
completely wet (CW) or partially wet (PW). Most experiments on SDSD have been performed 
on polymer blends, fluid mixtures, etc., where hydrodynamic effects play an important
role in the intermediate and late stages of phase separation. However, there have been very few numerical
investigations of SDSD with hydrodynamics. 

We undertook comprehensive molecular dynamics (MD) simulations to study the kinetics 
of SDSD in this paper. The MD simulations are performed with a Nos\'{e}-Hoover thermostat,
which naturally incorporates hydrodynamic effects.
In both CW and PW cases, the surface becomes the origin of SDSD 
waves, which propagate into the bulk. The typical SDSD profile consists of a 
multi-layered morphology, i.e., a wetting layer followed by a depletion layer, etc. We 
are interested in understanding the role of hydrodynamics in driving the growth of the bulk 
domain size and the wetting layer. At early times, the wetting layer grows diffusively 
with time ($R_1\sim t^{1/3}$). However, there is a crossover to a convective regime, and the 
late-stage dynamics is $R_1\sim t$. There is also a corresponding crossover in 
the growth dynamics of the bulk domain size $L(t)$. Due to computational limitations, our
MD simulations are as yet unable to access the inertial hydrodynamic regime (with
$L, R_1 \sim t^{2/3}$) in either the bulk or the wetting-layer kinetics.

Our findings have significant implications for experiments on SDSD, as many of these 
are performed on fluid mixtures. We hope that these results will provoke fresh 
experimental interest in this problem, and our theoretical results will be subjected 
to an experimental confirmation. 

\vspace*{0.2cm}
{\noindent \bf Acknowledgments} \\

PKJ acknowledges the University Grants Commission, India for financial support.

\newpage

\end{document}